\documentclass[pra,a4paper,aps,twocolumn,superscriptaddress,longbibliography]{revtex4-2}

\usepackage{amsmath, empheq,braket,amsthm,amssymb,graphics,amsfonts,array,color,subfigure, dsfont, graphicx, physics, bm}

\definecolor{myurlcolor}{rgb}{0,0,0.7}
\definecolor{myurlcolor1}{rgb}{0,0.7,0.1}
\definecolor{myrefcolor}{rgb}{0,0,0.7}

\usepackage[unicode=true,pdfusetitle, bookmarks=true,bookmarksnumbered=false,bookmarksopen=false, breaklinks=false,pdfborder={0 0 0},backref=false,colorlinks=true, linkcolor=myrefcolor,citecolor=myurlcolor1,urlcolor=myurlcolor]{hyperref}
\usepackage{lipsum,MnSymbol}
\usepackage[utf8]{inputenc}  
\usepackage[T1]{fontenc}

\begin{document}
	
\title{Majorization ladder in bosonic Gaussian channels}

\author{Zacharie Van Herstraeten}
\affiliation{Centre for Quantum Information and Communication, \'Ecole polytechnique de Bruxelles, CP 165/59, Universit\'e libre de Bruxelles, 1050 Brussels, Belgium}
\affiliation{Wyant College of Optical Sciences, The University of Arizona, 1630 E. University Blvd., Tucson, AZ 85721}

\author{Michael G. Jabbour}
\affiliation{Department of Physics, Technical University of Denmark, 2800 Kongens Lyngby, Denmark}

\author{Nicolas J. Cerf}
\affiliation{Centre for Quantum Information and Communication, \'Ecole polytechnique de Bruxelles, CP 165/59, Universit\'e libre de Bruxelles, 1050 Brussels, Belgium}
\affiliation{Wyant College of Optical Sciences, The University of Arizona, 1630 E. University Blvd., Tucson, AZ 85721}

\begin{abstract}
	We show the existence of a majorization ladder in bosonic Gaussian channels, that is, we prove that the channel output resulting from the $n\text{th}$ energy eigenstate (Fock state) majorizes the channel output resulting from the $(n\!+\!1)\text{th}$ energy eigenstate (Fock state). This reflects a remarkable link between the energy at the input of the channel and a disorder relation  at its output as captured by majorization theory. This result was previously known in the special cases of a pure-loss channel and quantum-limited amplifier, and we achieve here its nontrivial generalization to any single-mode phase-covariant (or -contravariant) bosonic Gaussian channel. The key to our proof is the explicit construction of a column-stochastic matrix that relates the outputs of the channel for any two subsequent Fock states at its input. This is made possible by exploiting a recently found recurrence relation on multiphoton transition probabilities for Gaussian unitaries [M. G. Jabbour and N. J. Cerf, Phys. Rev. Research 3, 043065 (2021)]. Possible generalizations  and implications of these results are then discussed.\footnote{Invited contribution to \textit{Jonathan P. Dowling Memorial Special Issue: The Second Quantum Revolution}, edited by Zixin Huang and Pieter Kok, Special Topic Collection, AVS Quantum Science.}
\end{abstract}

\maketitle

\section{Introduction}


When setting up a communication system through any channel, it is natural to aim for the fastest and most reliable protocol that is allowed by the intrinsic physical properties of the channel.
In that respect, the ultimate description of a channel is provided by quantum mechanics, while the optimal communication rate can be predicted by using the tools of information theory \cite{shannon1948mathematical}.
The combination of these two theories makes it possible to equip any quantum channel with a classical capacity, which gives a limit on the number of classical bits of information that can faithfully be conveyed per use of the channel.
The single-use classical capacity of a quantum channel $\mathcal{M}$ is obtained by maximizing the Holevo information, namely \cite{SchumacherWestmoreland1997,Holevo1998}
\begin{equation}
    C(\mathcal{M})
    =
    \max\limits_{\lbrace p_i,\hat{\rho}_i\rbrace}
    \Bigg[S\bigg(
    \mathcal{M}\bigg(
    \sum_i p_i\hat{\rho}_i
    \bigg)\bigg)
    -
    \sum_i
    p_i \, S\big(\mathcal{M}(\hat{\rho}_i)\big)\Bigg],
    \label{eq:Holevo}
\end{equation}
where $S(\hat{\rho})$ denotes the von Neumann entropy of the density operator $\hat{\rho}$ and $\lbrace p_i,\hat{\rho}_i\rbrace$ characterizes an ensemble of symbol states $\hat{\rho}_i$ with associated probabilities $p_i$.
In the process of evaluating $C(\mathcal{M})$, a crucial step is often to determine the input states that minimize the output entropy of the channel [i.e., the second term of the right-hand side of Eq. \eqref{eq:Holevo}] as these states can be chosen as symbol states in order to construct an efficient communication protocol, which may attain $C(\mathcal{M})$ in some cases.

In this paper, we focus on optical communication channels, e.g., optical fibers or free-space links, which are often adequately described by the model of bosonic Gaussian channels. The classical capacity of these channels was thoroughly investigated in Refs.
\cite{Holevo-bosonic-channels99,Holevo-Werner2001}, but the expression for $C(\mathcal{M})$ was pending on the knowledge of the minimum-output-entropy states. In the common case of single-mode phase-insensitive bosonic Gaussian channels (denoted simply as BGCs in what follows), it was conjectured very early on that coherent states minimize the output entropy \cite{giovannetti2004minimum}.
Yet, it took ten years to find a proof of this conjecture, which was needed to ascertain the conjectured expression of the capacity of BGCs \cite{giovannetti2014ultimate}. This result was shortly followed by an even stronger statement, namely that the outputs resulting from coherent states majorize every other outputs \cite{mari2014quantum}.
A majorization relation is arguably the most fundamental way to express that a quantum state is more statistically disordered than another one.
Recall that a state $\hat{\rho}$ is said to majorize a state $\hat{\sigma}$ (written $\hat{\rho}\succ\hat{\sigma}$) if and only if its vector of eigenvalues obeys the corresponding majorization relation,
\begin{equation}
	\mathbf{p}\succ\mathbf{q}
	\qquad\Leftrightarrow\qquad
	\sum\limits_{k=0}^{n}
	p^\downarrow_k
	\geq
	\sum\limits_{k=0}^{n}
	q^\downarrow_k ,
	\quad
	\forall n ,
\end{equation}
where $\mathbf{p}$ (resp. $\mathbf{q}$) is the vector of eigenvalues of $\hat{\rho}$ (resp. $\hat{\sigma}$), and $p^\downarrow_k$ (resp. $q^\downarrow_k$) is the $k^\text{th}$ highest component of $\mathbf{p}$ (resp. $\mathbf{q}$).
The relation $\hat{\rho}\succ\hat{\sigma}$ implies in turn inequalities over the broad set of Schur-convex (-concave) functions, notably it implies that $S(\hat{\rho})\leq S(\hat{\sigma})$ as a consequence of the Schur-concavity of the von Neumann entropy $S$.

The importance of majorization relations in the context of BGCs was first suggested in Ref. \cite{Guha-PhD}. This motivated an early attempt at proving the minimum output entropy conjecture \cite{Garcia2012majorization}, which established  that the vacuum is the minimizer state over the restricted set of phase-invariant inputs (but lacking an argument that it is sufficient to consider phase-invariant minimizer states). Interestingly, this incomplete proof in \cite{Garcia2012majorization} also yielded the following side-result: the quantum-limited amplifier (denoted as $\mathcal{A}_g$, where $g$ is the gain) exhibits a majorization relation at the output for successive Fock states at its input, that is, $\mathcal{A}_g(\ket{i}\bra{i})\succ\mathcal{A}_g(\ket{i+1}\bra{i+1})$, $\forall i \in \mathbb{N}$.
A very similar result was later proven for the pure-loss channel (denoted as $\mathcal{E}_\eta$, with $\eta$ being the transmittance), that is, $\mathcal{E}_{\eta}(\ket{i}\bra{i})\succ\mathcal{E}_{\eta}(\ket{i+1}\bra{i+1})$, $\forall i \in \mathbb{N}$ \cite{gagatsos2013majorization}.

\begin{figure}[t!]
$
\begin{aligned}
	\includegraphics[width=0.67\linewidth]{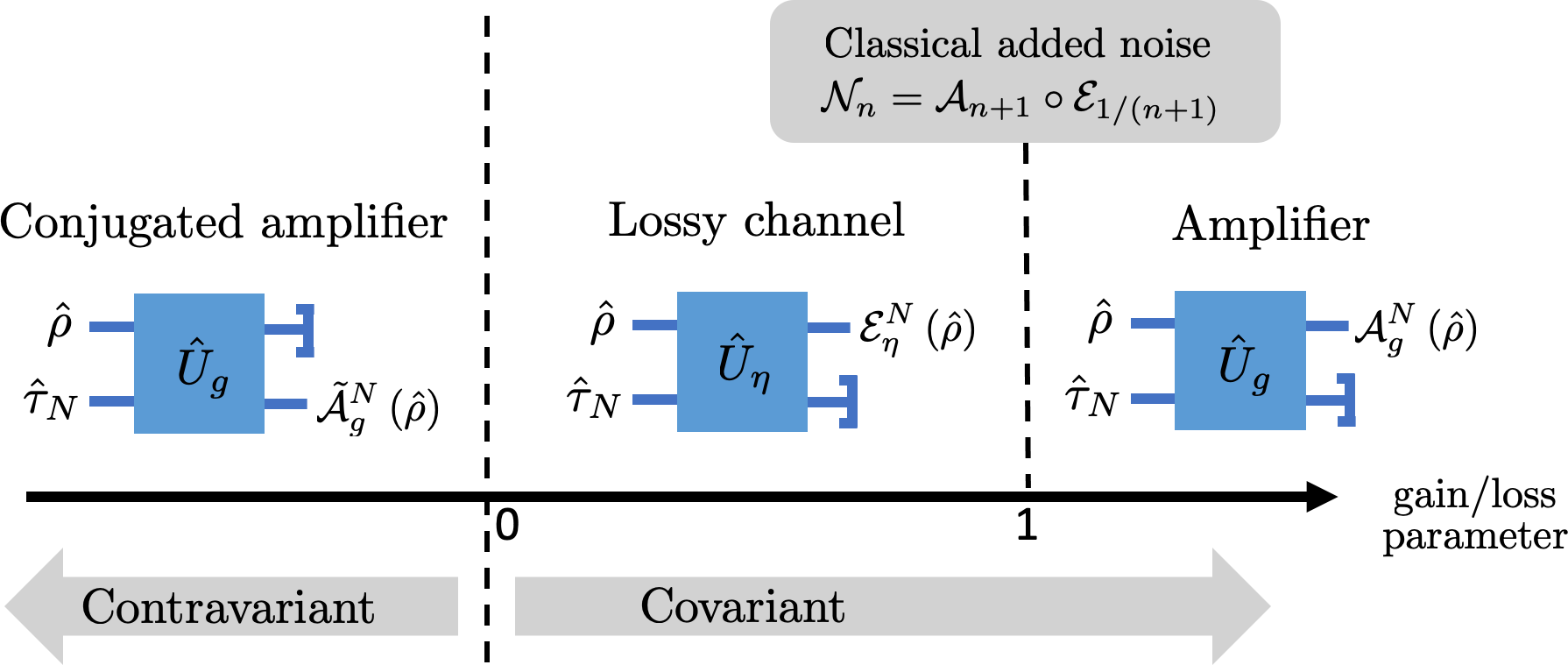}
\end{aligned}
$
	\caption{
		Optical BGCs are achievable with a beam-splitter, a two-mode squeezer, or a concatenation of both. Recall that 
		a beam-splitter of transmittance $\eta=\cos^2\theta$ is described by the unitary $\smash{\hat{U}_\eta=\exp(\theta(\hat{a}^\dagger\hat{b}-\hat{a}\hat{b}^\dagger))}$, while a two-mode squeezer of gain $g=\cosh^2 r$ is described by the unitary $\hat{U}_g=\exp(r(\hat{a}^\dagger \hat{b}^\dagger-\hat{a}\hat{b}))$.
		The lossy channel $\mathcal{E}^N_\eta$ is implemented with a beam-splitter acting on the input state $\hat{\rho}$ coupled with a thermal environmental state $\hat{\tau}_N$ of mean photon-number $N$; then the second output is discarded. Similarly, the amplifier $\mathcal{A}^N_g$ (resp. conjugated amplifier $\tilde{\mathcal{A}}^N_g$) is implemented as a two-mode squeezer acting on $\hat{\rho}\otimes\hat{\tau}_N$ and discarding the second output (resp. first output). Finally, the classical added noise channel $\mathcal{N}_n$ (with $n$ being the number of added thermal photons) is implemented as the concatenation of a pure-loss channel $\mathcal{E}_{1/(n+1)}$ followed by a quantum-limited amplifier $\mathcal{A}_{n+1}$.
		The lossy channel and amplifier are said to be phase-covariant and are, respectively, associated with a gain/loss parameter $\eta$ and $g$, while the conjugated amplifier is said to be  phase-contravariant and is associated with a gain/loss parameter $1-g$.
		The classical added noise is associated to a gain/loss parameter $1$.
	}
	\label{fig:bcg_summary}

\end{figure}

In practice, however, the pure-loss channel and quantum-limited amplifier represent only but a tiny fraction of the set of BGCs as they describe an ideal noiseless situation.
The presence of added noise can be taken into account by setting the environment of the Stinespring dilation of the channel to be a thermal state of mean photon-number $N$ (rather than a pure vacuum state), in which case channels $\mathcal{A}_g$ and $\mathcal{E}_\eta$ are denoted respectively as $\mathcal{A}^N_g$ and $\mathcal{E}^N_\eta$.
Supplemented with the classical additive noise channel $\mathcal{N}_n$, they span the entire set of covariant BGCs, while the contravariant BGCs are generated by the conjugated amplifier $\tilde{\mathcal{A}}^N_g$. 
We refer to Ref. \cite{giovannetti2014ultimate} for the detailed definitions of the above-mentioned channels, which are also summarized in Fig. \ref{fig:bcg_summary}.
It then comes as a natural question whether the output of Fock state $\ket{i}$ majorizes the output of the subsequent Fock state $\ket{i+1}$ in \textit{any} BGC.
Namely, does the relation
\begin{equation}
	\mathcal{M}(\ket{i}\bra{i})
	\succ
	\mathcal{M}(\ket{i+1}\bra{i+1}),
	\quad \forall i \in \mathbb{N}
	\label{eq:majorization_ladder}
\end{equation}
hold, where $\mathcal{M}$ stands for any of the channels $\mathcal{E}^N_\eta$, $\mathcal{A}^N_g$, $\mathcal{N}_n$ or $\tilde{\mathcal{A}}^N_g$?
We answer this question by the positive in the present paper. In the following, we will say that a quantum channel $\mathcal{M}$ obeys a \textit{majorization ladder} when it satisfies Eq. \eqref{eq:majorization_ladder}, so that
\begin{equation}
  \mathcal{M}(\ket{0}\bra{0})  \succ 
  \mathcal{M}(\ket{1}\bra{1})  \succ 
  \mathcal{M}(\ket{2}\bra{2})  \succ 
  \cdots \, ,
  	\label{eq:majorization_ladder_expanded}
\end{equation}
as a result of the transitivity of majorization. We also say that the vacuum state $\ket{0}$ is located at the bottom of the ladder.

Having in mind that any (covariant) BGC is equivalent to the concatenation of a quantum-limited amplifier and a pure-loss channel, it might at first sight seem an easy task to show that all (covariant) BGCs obey a majorization ladder (since both $\mathcal{A}_g$ and $\mathcal{E}_\eta$ do).
However, let us observe that the majorization ladder translates an energy relation at the input (Fock states are energy eigenstates) into a disorder relation at the output.
Proving the majorization ladder for the whole set of BGCs can then not be done via a simple concatenation argument.
Moreover, let us also stress that a majorization relation between two states is a strong statement (involving an infinite set of inequalities for states of a bosonic mode), and two states can in general be incomparable (in which case there is no majorization relation in either direction).
This makes the extension of the majorization ladder to all BGCs as presented here a strong and non-trivial result.

Our paper is structured as follows.
In Section \ref{sec:scheme}, we lay the general sketch of our proof, which relies on the construction of a column-stochastic matrix linking the eigenvalues of the output states associated with subsequent Fock states for any BGC. The matrix is built by exploiting a recently found recurrence relation on the transition probabilities for Gaussian unitaries \cite{jabbour2021multiparticle}.
In Section \ref{sec:gen_func_and_trans_prob}, we  use the generating function as a mathematical tool  in order to derive a recurrence relation satisfied by the output states of any BGC.
We then construct in Section \ref{sec:recurrence_matrix} the column-stochastic matrix which proves our main result.
We conclude in Section \ref{sec:conclusion} with some observations and further perspectives.

\onecolumngrid

\clearpage
\twocolumngrid

\section{Sketch of the proof}
\label{sec:scheme}

The first step of our proof is to use the fact that BGCs can be expressed via their Stinespring dilation as 
\begin{equation}
	\mathcal{M}(\hat{\rho})
	=
	\Tr_2\left[
	\hat{U}_G\left(
	\hat{\rho}\otimes\hat{\tau}_y
	\right)\hat{U}^\dagger_G
	\right],
	\label{eq:gaussian_channel}
\end{equation}
where  $\hat{\tau}_y=(1-y)\sum_{k=0}^{\infty} y^k\ket{k}\bra{k}$ is a thermal state of mean photon-number $N=y/(1-y)$, with $0\leq y<1$.
Here, the Gaussian unitary $\hat{U}_G$ is either $\hat{U}_\eta$ (for channel $\mathcal{E}^N_\eta$) or $\hat{U}_g$ (for channels $\mathcal{A}^N_g$ and $\tilde{\mathcal{A}}^N_g$), see Fig. \ref{fig:bcg_summary}.
Note that in the case of the conjugated amplifier $\tilde{\mathcal{A}}^N_g$, the partial trace is performed over mode $1$.
Although the additive noise channel $\mathcal{N}_n$ cannot be expressed in the form of \eqref{eq:gaussian_channel},  it can be taken as a limiting case of a lossy channel $\mathcal{E}^N_\eta$ (with $\eta\rightarrow 1$ and $N\rightarrow\infty$), or equivalently of an amplifier channel $\mathcal{A}^N_g$ (with $g\rightarrow 1$ and $N\rightarrow\infty$). 
These limits are made precise in Section \ref{sec:gen_func_and_trans_prob}.

\begin{figure}[t]
	\includegraphics[width=0.8\linewidth]{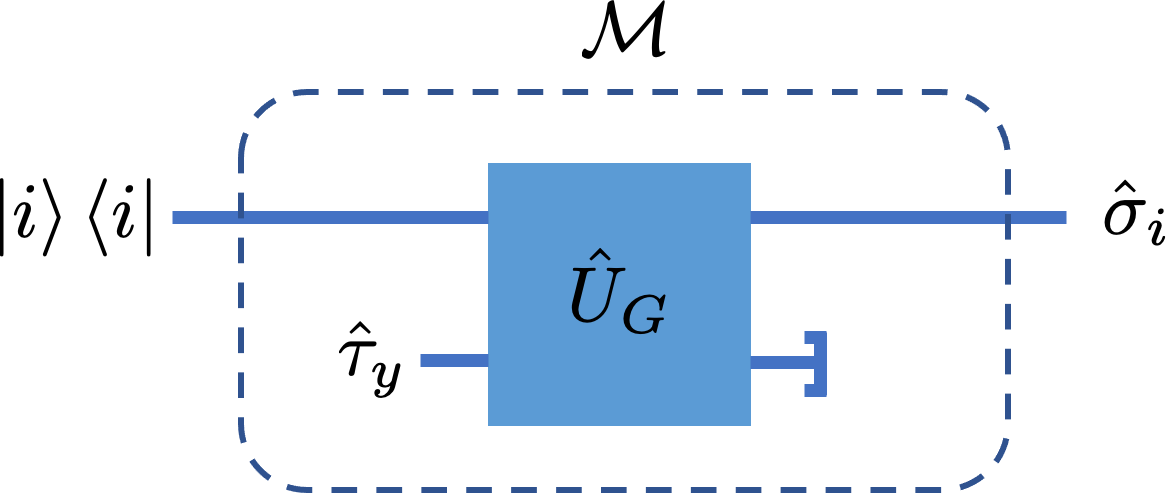}
	\caption{
		Stinespring dilation of a single-mode phase-covariant Gaussian channel $\mathcal{M}$. Here, the input state (Fock state $\ket{i}$) is coupled via the unitary $\hat{U}_G$ to a thermal state $\hat{\tau}_y$, resulting in the output state $\hat{\sigma}^{(i)}$ when tracing over the environment. The unitary $\hat{U}_G$ can be chosen to be the unitary $\hat{U}_\eta$ of a beam-splitter of transmittance $\eta$, or $\hat{U}_g$ for a two-mode squeezer of gain $g$.
	}
	\label{fig:channel_sigma_y}
\end{figure}

Let us define $\hat{\sigma}^{(i)}$ as the output of the $i^{\text{th}}$ Fock state in the channel $\mathcal{M}$ defined in \eqref{eq:gaussian_channel}, so that $\hat{\sigma}^{(i)}=\mathcal{M}(\ket{i}\bra{i})$, see Figure \ref{fig:channel_sigma_y}.
Since we consider a phase-covariant or phase-contravariant channel $\mathcal{M}$, we can always express $\hat{\sigma}^{(i)}$ as a mixture of Fock states
\begin{equation}
	\hat{\sigma}^{(i)}
	=
	\sum\limits_{n=0}^{\infty}
	T^{(i)}_n
	\ket{n}\bra{n},
\end{equation}
where $T^{(i)}_n$ is nothing but the transition probability to measure $n$ photons at the output of $\mathcal{M}$ when $i$ photons have been sent at its input. Our objective is thus to prove the majorization ladder 
\begin{equation}
\hat{\sigma}^{(i)}\succ\hat{\sigma}^{(i+1)} , \qquad \forall i\in \mathbb{N}.
\label{eq:central-majorization}
\end{equation}
At this point, we may define the vector $\mathbf{t}^{(i)}$ as the vector with components $\left(\mathbf{t}^{(i)}\right)_n=T^{(i)}_n$, so that $\mathbf{t}^{(i)}$ is the vector of eigenvalues of $\hat{\sigma}^{(i)}$.
The majorization ladder \eqref{eq:central-majorization} is then equivalent to $\mathbf{t}^{(i)}\succ\mathbf{t}^{(i+1)}$.

A notable feature of majorization theory is that the relation $\mathbf{p}\succ\mathbf{q}$ is equivalent to the existence of a column-stochastic matrix $\mathbf{D}$ ($D_{ij}\geq 0$, $\sum_{i}D_{ij}=1$, $\sum_{j}D_{ij}\leq 1$) such that $\mathbf{q}=\mathbf{D}\, \mathbf{p}$ \cite{infinite-majorization,Garcia2012majorization}. Note the subtlety that although the columns sum to 1 (hence, the term column-stochastic), the 
sum of each row must be $\le 1$ for  infinite-dimensional vectors $\mathbf{t}^{(i)}$ (for finite-dimensional vectors, the rows must sum to 1, just as the columns, and the matrix associated with majorization is called doubly stochastic). 

In what follows, we are going to construct such a column-stochastic matrix $\mathbf{D}$ by computing the generating function of the transition probabilities $T^{(i)}_n$, which will enable us to derive a recurrence relation obeyed by $T^{(i)}_n$.
Using this recurrence relation recursively, we will construct a matrix linking the components of $\mathbf{t}^{(i+1)}$ to the components of $\mathbf{t}^{(i)}$ and show that this matrix is column-stochastic, concluding the proof.

\section{Generating function of transition probabilities}
\label{sec:gen_func_and_trans_prob}

The generating function of a sequence $\lbrace c_n\rbrace$ with $n\in\mathbb{N}$ is a function that encapsulates all information about the sequence and
is defined as
\begin{equation}
	f(z)=\sum\limits_{n=0}^{\infty} c_n \, z^n.
\end{equation}
The sequence $\lbrace c_n\rbrace$ can be retrieved from $f(z)$ by using $c_n=(1/n!)(\mathrm{d}^n f/\mathrm{d}z^n)\vert_{z=0}$.
The generating function obeys the shifting property, namely, multiplying it by the variable $z$ corresponds to a right shift of the sequence: if $f(z)\leftrightarrow\lbrace c_n\rbrace$, then $z\, f(z)\leftrightarrow\lbrace c_{n-1}\rbrace$, with the convention that $c_n=0$ for $n<0$. Along with the shifting property, we will also use the fact that a constant generating function is associated with an empty sequence where only $c_0$ is non-zero, that is, $f(z)=1\leftrightarrow\lbrace\delta_n\rbrace$.

\begin{figure}[t]
	\includegraphics[width=0.95\linewidth]{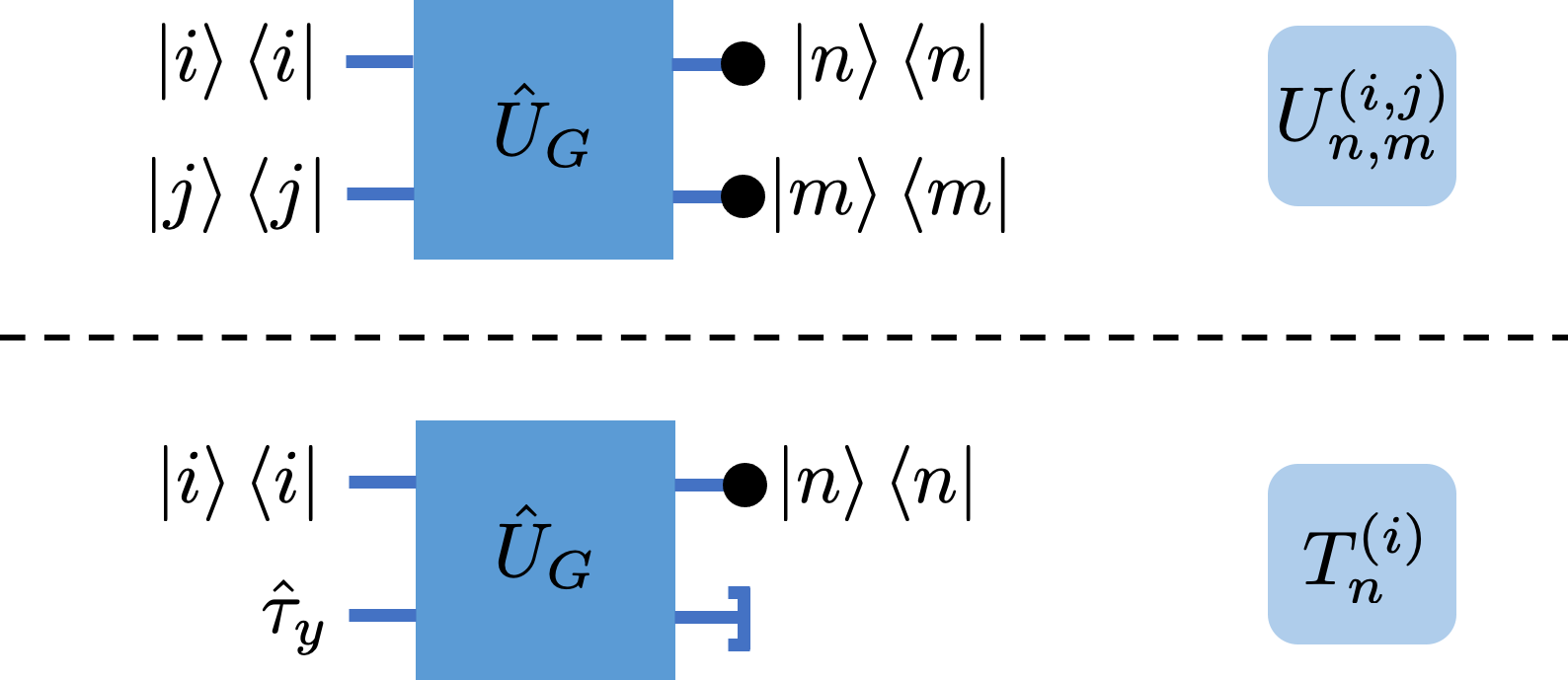}
	\caption{
		Transition probability $U^{(i,j)}_{n,m}$ for the Gaussian unitary $\hat{U}_G$ and transition probability $T^{(i)}_n$ for the bosonic Gaussian channel $\mathcal{M}$.
	}
	\label{fig:trans_probs}
\end{figure}

Let us define the transition probabilities of the Gaussian unitary $\hat{U}_G$ in the Fock basis as 
\begin{equation}
	U^{(i,j)}_{n,m}
	=
	\vert
	\bra{n,m}\hat{U}_G\ket{i,j}
	\vert^2
\end{equation}
with $i,j,n,m \in \mathbb{N}$, see Fig. \ref{fig:trans_probs}. The generating function associated with the sequence $\lbrace U^{(i,j)}_{n,m}\rbrace$, namely
\begin{equation}
	f(x,y,z,w)=
	\sum\limits_{i=0}^{\infty}
	x^i
	\sum\limits_{j=0}^{\infty}
	y^j
	\sum\limits_{n=0}^{\infty}
	z^n
	\sum\limits_{m=0}^{\infty}
	w^m
	\ 
	U^{(i,j)}_{n, m}  \, ,
\end{equation}
has a closed analytical expression when $\hat{U}_G$ is chosen to be the unitary of a beam-splitter or two-mode squeezer, see Ref.  \cite{jabbour2021multiparticle}.
It should be noted here that multi-index sequences are simply associated with multivariate generating functions, with one variable associated with each index.
Now, the generating function of the sequence $\lbrace T^{(i)}_n\rbrace$, namely,
\begin{equation}
	h(x,z)=
	\sum\limits_{i=0}^{\infty}
	x^i
	\sum\limits_{n=0}^{\infty}
	z^n
	\ 
	T^{(i)}_{n},
\end{equation}
can be obtained from the generating function of  $\lbrace U^{(i,j)}_{n,m}\rbrace$ as follows:
\begin{equation}
\label{eq-generating-function-h}
	\begin{split}
		h(x,z)
		&=
		\sum\limits_{i=0}^{\infty}x^i
		\sum\limits_{n=0}^{\infty}z^n
		\\
		&\quad \times
		\Tr
		\left[
		\hat{U}_G
		\left(
		\ket{i}\bra{i}
		\otimes
		\hat{\tau}_y
		\right)
		\hat{U}_G^\dagger
		\left(
		\ket{n}\bra{n}
		\otimes
		\hat{\mathds{1}}
		\right)
		\right]
		\\
		&=
		(1-y) \,
		\sum\limits_{i=0}^{\infty}x^i
		\sum\limits_{j=0}^{\infty}y^j
		\sum\limits_{n=0}^{\infty}z^n
		\sum\limits_{m=0}^{\infty} 1^m
		\\
		&\quad\times
		\Tr
		\left[
		\hat{U}_G
		\left(
		\ket{i}\bra{i}
		\otimes
		\ket{j}\bra{j}
		\right)
		\hat{U}_G^\dagger
		\left(
		\ket{n}\bra{n}
		\otimes
		\ket{m}\bra{m}
		\right)
		\right]
		\\
		&=
		(1-y) \,
		\sum\limits_{i=0}^{\infty}x^i
		\sum\limits_{j=0}^{\infty}y^j
		\sum\limits_{n=0}^{\infty}z^n
		\sum\limits_{m=0}^{\infty} 1^m
		\ 
		U^{(i,j)}_{n,m}
		\\
		&=
		(1-y) \,
		f(x,y,z,1)
	\end{split}
\end{equation}
In principle, the transition probabilities $T^{(i)}_n$ as pictured in Fig. \ref{fig:trans_probs} can thus be retrieved from  $h(x,z)$. Expression~\eqref{eq-generating-function-h} holds for both the noisy loss channel $\mathcal{E}^N_\eta$ and noisy amplifier  $\mathcal{A}^N_g$ provided $f(x,y,z,1)$ is replaced by its appropriate expression as found in Ref. \cite{jabbour2021multiparticle}. In the case of the conjugated amplifier $\tilde{\mathcal{A}}^N_g$, the partial trace is done over the first mode, so that the generating function is $h(x,z)=(1-y)\, f(x,y,1,z)$.
For the classical additive noise channel $\mathcal{N}_n$, the generating function is obtained as a limiting case of the one for  $\mathcal{E}^N_\eta$ with $\eta\rightarrow 1^-$ and $N\rightarrow\infty$, or equivalently of $\mathcal{A}^N_g$ with $g\rightarrow 1^+$ and $N\rightarrow\infty$.
The limit is carried out by setting, respectively, $(1-\eta)\, N=n$ or $(g-1)\, N=n$. Remarkably, the expression of $h(x,z)$ has a similar structure for the four channels $\mathcal{E}^N_\eta$, $\mathcal{A}^N_g$, $\mathcal{N}_n$, and $\tilde{\mathcal{A}}^N_g$,
and reads as
\begin{equation}
	h(x,z)
	=
	\frac{\chi}{1-\alpha \, x-\beta \, z -\gamma \, xz},
	\label{eq:generating_function_fraction}
\end{equation}
where the parameters $\alpha,\beta,\gamma$, and $\chi$ are defined in Table~\ref{table:coefficients}. Note that these parameters are constrained as a result of the trace-preservation condition of the channel, namely,
\begin{equation}
    \sum_{n=0}^\infty T^{(i)}_n = 1, \qquad \forall i,
\end{equation}
which translates into
\begin{equation}
\label{eq:constraint-on-h}
    h(x,1) = (1-x)^{-1}, \qquad \forall x.
\end{equation}
It is easy to check that Eq. \eqref{eq:constraint-on-h} implies the conditions
\begin{equation}
\label{eq:condition}
\alpha+\beta+\gamma = 1 , \qquad
\beta + \chi =1 ,
\end{equation}
so BGCs only depend on two independent parameters.

\renewcommand{\arraystretch}{2.0}
\begin{table}[t]
	\resizebox{0.8\columnwidth}{!}{%
	\begin{tabular}{l|c|c|c|c|}
		\cline{2-5}
		\multicolumn{1}{c|}{}                           & $\alpha$                  & $\beta$                      & $\gamma$                  & $\chi$                 \\ \hline
		\multicolumn{1}{|l|}{$\mathcal{E}^N_\eta$}      & $\frac{1-\eta}{1-\eta y}$ & $\frac{y(1-\eta)}{1-\eta y}$ & $\frac{\eta-y}{1-\eta y}$ & $\frac{1-y}{1-\eta y}$ \\ \hline
		\multicolumn{1}{|l|}{$\mathcal{A}^N_g$}         & $\frac{y(g-1)}{g-y}$      & $\frac{g-1}{g-y}$            & $\frac{1-gy}{g-y}$        & $\frac{1-y}{g-y}$      \\ \hline
		\multicolumn{1}{|l|}{$\mathcal{N}_n$}           & $\frac{n}{n+1}$           & $\frac{n}{n+1}$              & $\frac{1-n}{n+1}$         & $\frac{1}{n+1}$        \\ \hline
		\multicolumn{1}{|l|}{$\tilde{\mathcal{A}}^N_g$} & $\frac{gy-y+1}{g}$        & $\frac{g+y-1}{g}$            & $-y$                      & $\frac{1-y}{g}$        \\ \hline
	\end{tabular}
	}
\caption{Value of the parameters $\alpha,\beta,\gamma$, and $\chi$ for the four families of BGCs [see Eqs. \eqref{eq:generating_function_fraction} and \eqref{eq:recurrence_relation}]. Here, $\eta$ stands for the transmittance and $g$ for the gain, while $y$ is related to the mean photon-number $N$ of the environmental thermal state as $y=N/(N+1)$. The classical additive noise channel is defined as $\mathcal{N}_n=\mathcal{A}_{n+1}\circ\mathcal{E}_{1/(n+1)}$, where $n$ is the number of added thermal photons.}
\label{table:coefficients}
\end{table}

\section{From a recurrence relation to a column-stochastic matrix}
\label{sec:recurrence_matrix}

Equation \eqref{eq:generating_function_fraction} may be reexpressed as
\begin{equation}
	h(x,z)=\alpha \, x\, h(x,z)+\beta \, z \, h(x,z)+\gamma \, xz \, h(x,z)+\chi 
	\label{eq:gen_func_recurence_relation}
\end{equation}
which can be converted into the following recurrence relation on the transition probabilities,
\begin{equation}
	T_n^{(i)}
	=
	\alpha \, T_n^{(i-1)}
	+\beta \, T^{(i)}_{n-1}
	+\gamma \, T^{(i-1)}_{n-1}
	+\chi \, \delta^{(i)}_n  ,
	\label{eq:recurrence_relation}
\end{equation}
where we have used the shifting property, namely,
\begin{equation}
	\begin{split}
		h(x,z) 
		&\quad\longleftrightarrow\quad 
		\left\lbrace T_n^{(i)}\right\rbrace,
		\\
		x\ h(x,z) 
		&\quad\longleftrightarrow\quad
		\left\lbrace T_n^{(i-1)}\right\rbrace,
		\\
		z\ h(x,z)
		&\quad\longleftrightarrow\quad
		\left\lbrace T_{n-1}^{(i)}\right\rbrace,
		\\
		xz\ h(x,z)
		&\quad\longleftrightarrow\quad
		\left\lbrace T_{n-1}^{(i-1)}\right\rbrace.
	\end{split}
\end{equation}
Remember that the parameters $\alpha,\beta,\gamma$, and $\chi$ depend on the considered channel and are given in Table \ref{table:coefficients}.

By definition, $\delta^{(i)}_n$ is non-zero if and only if $i=n=0$, so that $\chi=T^{(0)}_0$ is the probability to measure zero photon at the output for a vacuum input. For any other component $T^{(i)}_n$, we have either $i>0$ or $n>0$, so that we can omit the last term of the right-hand side of Eq. \eqref{eq:recurrence_relation}.
Our goal now is to link the components of vector $\mathbf{t}^{(i)}$ to those of vector $\mathbf{t}^{(i-1)}$, so that in the right-hand side of Eq. \eqref{eq:recurrence_relation} we should have terms with index $i-1$ only (and get rid of the term in $T^{(i)}_{n-1}$). In order to do so, we recursively apply the recurrence relation as follows:
\begin{equation}
	\begin{split}
		T_n^{(i)} 
		&=
		\alpha T_n^{(i-1)}+\left( \gamma T_{n-1}^{(i-1)} +\beta T_{n-1}^{(i)} \right)
		\\[1.5em]
		&=
		\alpha T_n^{(i-1)}+\gamma T_{n-1}^{(i-1)} +\beta 
		\left(
		\alpha T_{n-1}^{(i-1)}+\gamma T_{n-2}^{(i-1)} +\beta T_{n-2}^{(i)}
		\right)
		\\[1.5em]
		&=
		\alpha T_n^{(i-1)}+(\gamma+\beta\alpha) T_{n-1}^{(i-1)} 
		+\beta \left(\gamma T_{n-2}^{(i-1)}
		+\beta T_{n-2}^{(i)} \right)
		\\[1.5em]
		&=
		\alpha T_n^{(i-1)}+(\gamma+\beta\alpha) T_{n-1}^{(i-1)} 
		+\beta\gamma T_{n-2}^{(i-1)}
		\\
		&
		\qquad\qquad\qquad
		+\beta^2 
		\left(
		\alpha T_{n-2}^{(i-1)}+\gamma T_{n-3}^{(i-1)} +\beta T_{n-3}^{(i)}
		\right)
		\\[1.5em]
		&=
		\alpha T_n^{(i-1)}
		+(\gamma+\beta\alpha) T_{n-1}^{(i-1)} 
		+\beta(\gamma+\beta\alpha) T_{n-2}^{(i-1)}
		\\
		&
		\qquad\qquad\qquad
		+\beta^2 \left(\gamma T_{n-3}^{(i-1)} 
		+\beta T_{n-3}^{(i)} \right)
		\\[1.5em]
		&=
		\cdots
	\end{split}
\end{equation}
By induction, this results into the closed expression
\begin{equation}
    T_n^{(i)} = \alpha \, T^{(i-1)}_n
		+
		\nu \, 
		\sum\limits_{k=1}^{n}
		\beta^{k-1} \, 
		T^{(i-1)}_{n-k} \, ,
\end{equation}
where we have defined the parameter $\nu=\gamma+\beta\alpha$. This links each component of vector $\mathbf{t}^{(i)}$ to the components of vector $\mathbf{t}^{(i-1)}$, as we wanted. In matrix terms, we thus have
\begin{equation}
	\underbrace{
		\begin{pmatrix}
			T_0^{(i)}
			\\
			T_1^{(i)}
			\\
			T_2^{(i)}
			\\
			T_3^{(i)}
			\\
			T_4^{(i)}
			\\
			\vdots
		\end{pmatrix}
	}_{\displaystyle \mathbf{t}^{(i)}}
	=
	\underbrace{
		\begin{pmatrix}
			\alpha & 0 & 0 & 0 & 0 &  \cdots 
			\\
			\nu & \alpha & 0 & 0 & 0 &  \cdots 
			\\
			\nu\beta & \nu & \alpha & 0 & 0 &  \cdots 
			\\
			\nu\beta^2 & \nu\beta & \nu & \alpha & 0 &  \cdots
			\\
			\nu\beta^3 & \nu\beta^2 & \nu\beta & \nu & \alpha & \cdots
			\\
			\vdots & \vdots & \vdots & \vdots & \vdots &  \ddots
		\end{pmatrix}
	}_{\displaystyle \mathbf{D}}
	\underbrace{
		\begin{pmatrix}
			T_0^{(i-1)}
			\\
			T_1^{(i-1)}
			\\
			T_2^{(i-1)}
			\\
			T_3^{(i-1)}
			\\
			T_4^{(i-1)}
			\\
			\vdots
		\end{pmatrix}
	}_{\displaystyle \mathbf{t}^{(i-1)}}
\end{equation}
with the entries of matrix $\mathbf{D}$ being given by
\begin{equation}
	D_{kl}
	=
	\alpha\, \delta(k-l)
	+
	\nu\, \beta^{k-l-1}\, \Theta(k-l-1),
\label{eq:entries-of-D}	
\end{equation}
where $\delta(z)=1$ if $z=0$ and is zero otherwise, and $\Theta(z)=1$ if $z\geq 0$ and is zero otherwise.
The last step of our proof is to check that $\mathbf{D}$ is indeed column-stochastic (this has to be done separately for the four channels).
Given the values of Table \ref{table:coefficients}, it is straightforward to show that $\alpha\geq 0$, $\beta\geq 0$ and $\nu\geq 0$, so all entries of $\mathbf{D}$ are non-negative.
Notice also that the sum of each column can be easily computed (since $\beta<1$) as
\begin{equation}
	\alpha+\nu\sum\limits_{k=0}^{\infty}\beta^k=
	\frac{\alpha+\gamma}{1-\beta} 
\end{equation}
which is equal to 1 as a result of Eq. \eqref{eq:condition}. The key point is to notice, from the structure of matrix $\mathbf{D}$, that its rows are simply truncated columns (with the remaining entries being pasted with zeros), which implies that the sum of each row is less than or equal to $1$. From this, we conclude that $\mathbf{D}$ is column-stochastic in our relation $\mathbf{t}^{(i)} = \mathbf{D} \,
\mathbf{t}^{(i-1)}$, which completes the proof of Eq. \eqref{eq:central-majorization}.

It should be noted that the column-stochastic matrix linking two vectors satisfying a majorization relation is, in general, not unique. Remarkably, the elements of the matrix $\mathbf{D}$ that we have found in Eq. \eqref{eq:entries-of-D} do not depend on the index $i$. Hence, the matrix $\mathbf{D}$ is only a function of the considered channel, which implies the simple relation
\begin{equation}
	\mathbf{t}^{(i)}
	=
	\mathbf{D}^k \, 
	\mathbf{t}^{(i-k)}
	=
	\mathbf{D}^i \,
	\mathbf{t}^{(0)}
	\label{eq:invariance_matrix_D}
\end{equation}
where $k$ can be chosen from $0$ to $i$. Note that any power of a column-stochastic matrix remains column-stochastic, so that Eq. \eqref{eq:invariance_matrix_D} directly implies
the majorization relations 
\begin{equation}
	 \mathcal{M}(\ket{i}\bra{i})
	\succ
	\mathcal{M}(\ket{i-k}\bra{i-k})
	\succ
	\mathcal{M}(\ket{0}\bra{0})
    \label{eq:invariance_matrix_D_majoriz}
\end{equation}
where $0\le k\le i$, which can also be understood as a consequence of the transitivity of majorization.

\section{Discussion and perspectives}
\label{sec:conclusion}

We have shown that all (phase-covariant or phase-contravariant) BGCs satisfy an intrinsic ladder of majorization relations \eqref{eq:majorization_ladder_expanded}, which connects the energy eigenstates at the input of the channel to the disorder of its output. Specifically, any two Fock states $\ket{i}$ and $\ket{j}$ result into output states that satisfy the majorization relation
\begin{equation}
	 \mathcal{M}(\ket{i}\bra{i})
	\succ
	\mathcal{M}(\ket{j}\bra{j})
	\quad \mathrm{if~~} i\le j.
\label{eq:majoriz_i_j}
\end{equation}
A major consequence of Eq. \eqref{eq:majoriz_i_j} is that we can 
relate the output von Neumann entropy produced by any two Fock states in any BGC as follows,
\begin{equation}
	 S\big(\mathcal{M}(\ket{i}\bra{i})\big)
	\le
	S\big(\mathcal{M}(\ket{j}\bra{j})\big)
	\quad \mathrm{if~~} i\le j  ,
\end{equation}
which also implies a monotonous entropy chain
\begin{equation}
	 S\big(\mathcal{M}(\ket{0}\bra{0})\big)
	\le
	S\big(\mathcal{M}(\ket{1}\bra{1})\big)
	\le
	S\big(\mathcal{M}(\ket{2}\bra{2})\big)
    \le  \cdots
\end{equation}
This can of course also be extended to all Schur-concave functions, including all Rényi entropies.

It is tempting to try generalizing the majorization ladder \eqref{eq:majorization_ladder_expanded} and assume a direct connection between the input energy and output disorder. Namely, one could expect that if $\hat{\rho}$ and $\hat{\sigma}$ are any two states such that the energy of $\hat{\rho}$ is lower than the energy of $\hat{\sigma}$, then the corresponding outputs satisfy the majorization relation $\mathcal{M}(\hat{\rho})
	\succ
	\mathcal{M}(\hat{\sigma})$.
It is easy, however, to find a counterexample. More subtly, we may want to replace the energy comparison at the input with a so-called Fock majorization relation as defined in Ref. \cite{Jabbour_2016}.  Fock majorization (denoted as $\succ_F$) corresponds to ``unordered'' majorization in Fock basis and is such that $\hat{\rho}\succ_F \hat{\sigma}$ if the energy of $\hat{\rho}$ is lower than the energy of $\hat{\sigma}$ (the converse does not hold).
It is trivial to see that $\ket{i}\bra{i}\succ_F \ket{j}\bra{j}$ if $i\le j$. A natural extension of the majorization ladder could then be that if $\hat{\rho}$ and $\hat{\sigma}$ are any two states that satisfy $\hat{\rho}\succ_F \hat{\sigma}$, then 
$\mathcal{M}(\hat{\rho}) \succ \mathcal{M}(\hat{\sigma})$. Unfortunately, this statement can be proven wrong too. Instead, what we know is that the Fock-majorization relation is preserved across any BGC \cite{Jabbour_2016}, that is, $\hat{\rho}\succ_F \hat{\sigma}$ implies $\mathcal{M}(\hat{\rho}) \succ_F \mathcal{M}(\hat{\sigma})$, where the output states must be compared using Fock majorization too. 
According to Ref. \cite{Jabbour_2016}, this Fock-majorization preservation is equivalent to the existence of a Fock-majorization ladder,  namely,
\begin{equation}
    \mathcal{M}(\ket{i}\bra{i})
    \succ_{F}
    \mathcal{M}(\ket{j}\bra{j})
	\quad \mathrm{if~~} i\le j ,
\end{equation}
which also appears as a straightforward consequence of the structure of the matrix $\mathbf{D}$ (it is lower-triangular). 
It thus seems that the majorization ladder 	\eqref{eq:majorization_ladder_expanded} that we have found in this paper is the strongest result that can be obtained along this line.

At the heart of this majorization ladder, we have found a closed expression for the column-stochastic matrix $\mathbf{D}$ for the entire set of BGCs, which provides a recurrence relation on the transition probabilities $T_n^{(i)}$. It should be stressed that the derivation of $\mathbf{D}$ is far from obvious because it requires the evaluation of non-Gaussian matrix elements of Gaussian unitaries in the Fock basis. This is where the generating function comes into play because it enables expressing these matrix elements by using the Gaussian toolbox only \cite{jabbour2021multiparticle}. Actually, the trick is not to access these matrix elements individually, but to exploit the recurrence equation that they obey (as implied by the generating function) in order to prove the majorization ladder \eqref{eq:majorization_ladder_expanded}.
The latter is expected to yield a valuable ingredient for understanding new properties of BGCs. Below, we exhibit properties that can either be deduced from the majorization ladder or that can be conjectured based on numerical evidence.

\medskip
(1) Consider two input states $\hat{\rho}$ and $\hat{\sigma}$, which are statistical mixtures of Fock states such that $\hat{\sigma}$ is shifted by $k$ steps with respect to $\hat{\rho}$, namely
\begin{equation}
	\hat{\rho}
	=
	\sum\limits_{i=0}^{\infty}
	c_{i}\ket{i}\bra{i},
	\qquad
	\hat{\sigma}
	=
	\sum\limits_{i=0}^{\infty}
	c_{i}\ket{i+k}\bra{i+k}
	\label{eq:shifted_mixture}
\end{equation}
with $\lbrace c_i\rbrace$ being a probability distribution and $k\ge 0$. Since $\hat{\rho}$ and $\hat{\sigma}$ are mixtures, it seems difficult at first sight to conclude on a majorization relation at the output 
\begin{equation}
    \mathcal{M}(\hat{\rho})\succ\mathcal{M}(\hat{\sigma})
    \label{eq:majorization-to-be-confirmed}
\end{equation}
even though each individual term of the two mixtures satisfies $\mathcal{M}(\ket{i}\bra{i})\succ\mathcal{M}(\ket{i+k}\bra{i+k})$. Indeed, majorization relations are not conserved under convex mixing: starting from two sets of vectors $\lbrace\mathbf{x}_i\rbrace$ and $\lbrace\mathbf{y}_i\rbrace$ such that $\mathbf{x}_i\succ\mathbf{y}_i$, $\forall i$, it is not true to infer that $\sum_i c_i\mathbf{x}_i\succ\sum_i c_i\mathbf{y}_i$ for any probability distributions $\lbrace c_i\rbrace$.
However, using Eq.~\eqref{eq:invariance_matrix_D} and defining $\mathbf{p}$ and $\mathbf{q}$
such that $\mathcal{M}(\hat{\rho})=\sum_n p_n\ket{n}\bra{n}$ and $\mathcal{M}(\hat{\sigma})=\sum_n q_n\ket{n}\bra{n}$, it is easily shown that $\mathbf{q}=\mathbf{D}^k\mathbf{p}$, from which we confirm Eq. \eqref{eq:majorization-to-be-confirmed}. Again, the simplification arises from the observation that $\mathbf{D}$ does not depend on the index of the Fock state it is applied to (as well as the fact that any power of a column-stochastic matrix is also column-stochastic).

\medskip
(2) Now, consider the same state $\hat{\sigma}$ as in Eq. \eqref{eq:shifted_mixture} but let us compare its associated output $\mathcal{M}(\hat{\sigma})=\sum_n q_n\ket{n}\bra{n}$ with the output $\mathcal{M}(\ket{k}\bra{k})=\sum_n T^{(k)}_n\ket{n}\bra{n}$ associated with the $k^\text{th}$ Fock state.
A simple calculation shows that
\begin{align}
    \mathbf{q}
    =
    \sum\limits_{i=0}^{\infty}
    c_i
    \mathbf{t}^{(i+k)}
    =
    \left(
    \sum\limits_{i=0}^{\infty}
    c_i
    \mathbf{D}^i
    \right)
    \mathbf{t}^{(k)},
\end{align}
where we have used Eq. \eqref{eq:invariance_matrix_D}.
Notice that any convex mixture of column-stochastic matrices is itself column-stochastic, which implies  that $\mathbf{t}^{(k)}\succ \mathbf{q}$ or, equivalently, that 
\begin{equation}
\mathcal{M}(\ket{k}\bra{k})  \succ \mathcal{M}(\hat{\sigma}).
\end{equation}
In other words, the output associated with a mixture of Fock states is majorized by the output that is associated with the lowest Fock state contained in the mixture, which is a distinct result from Eq. \eqref{eq:majorization-to-be-confirmed}.

Beyond properties (1) and (2), let us mention another majorization ladder that we expect to hold in BGCs although it is only supported by numerical evidence. The majorization ladder \eqref{eq:majorization_ladder_expanded} can be viewed as a path from state $\ket{0}$ to state $\ket{k}$ such that the corresponding output states monotonically majorize pairwise along the path. This can be viewed as a refinement of the majorization relation of Ref. \cite{mari2014quantum}, which implies that any state (thus in particular $\ket{k}$) results in an output that is majorized by the output resulting from the vacuum state $\ket{0}$, that is, $\mathcal{M}(\ket{0}\bra{0})  \succ \mathcal{M}(\ket{k}\bra{k})$. Instead, the majorization ladder \eqref{eq:majorization_ladder_expanded} involves all the intermediate Fock states between  $\ket{0}$ and $\ket{k}$.
One may expect the existence of a similar path but replacing the vacuum state with a passive state at the bottom of the ladder. Consider a sequence of isospectral states, that is, mixed states having the same vector of eigenvalues (among them, the passive state is the one that has the lowest energy). It was shown in Ref.~\cite{depalma2016passive} that, for all BGCs,  any state $\hat{\rho}$ produces an output that is majorized by the output resulting from the passive state $\hat{\rho}^{\downarrow}$ that is isospectral with $\hat{\rho}$. It is tempting to look for a path of isospectral states at the input resulting in a majorization ladder at the output, just as in Eq. \eqref{eq:majorization_ladder_expanded}. We do not know how to do this in general, but have numerically observed the following special case: 

\medskip
(3) Consider a mixture of Fock states of the form
\begin{equation}
\hat{\rho}_{n_0,n_1,n_2,\cdots}
=\frac{1}{N} \left(n_0 \ket{0}\bra{0}+
n_1 \ket{1}\bra{1}+
n_2 \ket{2}\bra{2}+ \cdots \right)
\end{equation}
restricting to the special case where the numbers $n_i$'s can only take two values, 0 or 1, and $N=\sum_i n_i$ is the number of nonvanishing components. Starting, for example, from state
$\hat{\rho}_{101001}$, we can define a path that ends with the corresponding passive state $\hat{\rho}_{111000}$ and along which the output states monotonically majorize pairwise. This path is valid provided that the following (conjectured) property holds: if we keep the components of the lowest Fock states unchanged up to some level and then permute the upper components so to minimize the energy, the corresponding output states satisfy a majorization relation. This is better explained with an example, namely   
\begin{equation}
    \mathcal{M}(\hat{\rho}_{\cdots \, | \, 011}) \succ \mathcal{M}(\hat{\rho}_{\cdots \, | \, 110})
    \label{eq:conjecture}
\end{equation}
where the ``core'' binary sequence denoted as $\cdots$ is arbitrary, while the sequence $011$ is compared with its passive counterpart $110$ (where all one's have been brought to the left). By applying this conjectured property recursively starting for example from state $\hat{\rho}_{101001}$, one gets 
\begin{equation}
\begin{split}
    \mathcal{M}(\hat{\rho}_{1010 \, | \, 10}) \succ  \mathcal{M}(\hat{\rho}_{1010 \, | \, 01}) \\
    \mathcal{M}(\hat{\rho}_{101 \, | \, 100}) \succ \mathcal{M}(\hat{\rho}_{101 \, | \, 010})\\
    \mathcal{M}(\hat{\rho}_{1 \, | \, 11000}) \succ \mathcal{M}(\hat{\rho}_{1 \, | \, 01100})
\end{split}    
\end{equation}
where the vertical bars separate the ``core'' from the rest of the sequence. This yields the majorization ladder 
\begin{equation}
\mathcal{M}(\hat{\rho}_{111000}) \succ
    \mathcal{M}(\hat{\rho}_{101100}) \succ
    \mathcal{M}(\hat{\rho}_{101010})  \succ
        \mathcal{M}(\hat{\rho}_{101001}) 
\label{eq:natural-path}
\end{equation}
which holds for all BGCs conditionally on conjecture~\eqref{eq:conjecture}. We have numerically verified  this conjecture, but have not been able to prove it analytically. Further, we should note that it is not valid any more if the components $n_i$'s take other values than 0 or 1. Yet, Eq. \eqref{eq:natural-path} provides a neat construction for a path of input states of increasing energy producing output states of increasing disorder as measured by majorization theory.

Finally, we should note that all the results of this work are restricted to phase-insensitive Gaussian channels, where the environment is taken to be a thermal state. However, in the case of phase-sensitive Gaussian channels that are built with a squeezed thermal state in the environment, we should observe a similar majorization ladder with squeezed Fock states at the input (with the same squeezing parameter as the environment). 
Another interesting extension of the current work could be to investigate the existence of a majorization ladder (or majorization lattice) for multimode bosonic Gaussian channels that are gauge-covariant (i.e., covariant with respect to orthogonal rotations in phase space). We anticipate that this majorization-based approach holds the promise of shedding new light on the fundamental properties of quantum optical communication channels.

\begin{acknowledgements}
N.J.C. would like to thank Saikat Guha for introducing him to majorization theory in bosonic Gaussian channels and for enlightening discussions on this topic. Z.V.H. acknowledges support from the Belgian American Educational Foundation. M.G.J. acknowledges support from the Carlsberg Foundation under Grant CF19-0313. N.J.C. acknowledges support from the Fonds de la Recherche Scientifique (F.R.S.--FNRS)  under  Grant T.0224.18 and from the European Union under project ShoQC within the ERA-NET Cofund in Quantum Technologies (Quant\-ERA) program.
\end{acknowledgements}

\bibliography{majorchain}

\begin{thebibliography}{15}%
\makeatletter
\providecommand \@ifxundefined [1]{%
 \@ifx{#1\undefined}
}%
\providecommand \@ifnum [1]{%
 \ifnum #1\expandafter \@firstoftwo
 \else \expandafter \@secondoftwo
 \fi
}%
\providecommand \@ifx [1]{%
 \ifx #1\expandafter \@firstoftwo
 \else \expandafter \@secondoftwo
 \fi
}%
\providecommand \natexlab [1]{#1}%
\providecommand \enquote  [1]{``#1''}%
\providecommand \bibnamefont  [1]{#1}%
\providecommand \bibfnamefont [1]{#1}%
\providecommand \citenamefont [1]{#1}%
\providecommand \href@noop [0]{\@secondoftwo}%
\providecommand \href [0]{\begingroup \@sanitize@url \@href}%
\providecommand \@href[1]{\@@startlink{#1}\@@href}%
\providecommand \@@href[1]{\endgroup#1\@@endlink}%
\providecommand \@sanitize@url [0]{\catcode `\\12\catcode `\$12\catcode
  `\&12\catcode `\#12\catcode `\^12\catcode `\_12\catcode `\%12\relax}%
\providecommand \@@startlink[1]{}%
\providecommand \@@endlink[0]{}%
\providecommand \url  [0]{\begingroup\@sanitize@url \@url }%
\providecommand \@url [1]{\endgroup\@href {#1}{\urlprefix }}%
\providecommand \urlprefix  [0]{URL }%
\providecommand \Eprint [0]{\href }%
\providecommand \doibase [0]{https://doi.org/}%
\providecommand \selectlanguage [0]{\@gobble}%
\providecommand \bibinfo  [0]{\@secondoftwo}%
\providecommand \bibfield  [0]{\@secondoftwo}%
\providecommand \translation [1]{[#1]}%
\providecommand \BibitemOpen [0]{}%
\providecommand \bibitemStop [0]{}%
\providecommand \bibitemNoStop [0]{.\EOS\space}%
\providecommand \EOS [0]{\spacefactor3000\relax}%
\providecommand \BibitemShut  [1]{\csname bibitem#1\endcsname}%
\let\auto@bib@innerbib\@empty
\bibitem [{\citenamefont {Shannon}(1948)}]{shannon1948mathematical}%
  \BibitemOpen
  \bibfield  {author} {\bibinfo {author} {\bibfnamefont {C.~E.}\ \bibnamefont
  {Shannon}},\ }\bibfield  {title} {\bibinfo {title} {A mathematical theory of
  communication},\ }\href {https://doi.org/10.1002/j.1538-7305.1948.tb01338.x}
  {\bibfield  {journal} {\bibinfo  {journal} {The Bell System Technical
  Journal}\ }\textbf {\bibinfo {volume} {27}},\ \bibinfo {pages} {379}
  (\bibinfo {year} {1948})}\BibitemShut {NoStop}%
\bibitem [{\citenamefont {Schumacher}\ and\ \citenamefont
  {Westmoreland}(1997)}]{SchumacherWestmoreland1997}%
  \BibitemOpen
  \bibfield  {author} {\bibinfo {author} {\bibfnamefont {B.}~\bibnamefont
  {Schumacher}}\ and\ \bibinfo {author} {\bibfnamefont {M.~D.}\ \bibnamefont
  {Westmoreland}},\ }\bibfield  {title} {\bibinfo {title} {Sending classical
  information via noisy quantum channels},\ }\href
  {https://doi.org/10.1103/PhysRevA.56.131} {\bibfield  {journal} {\bibinfo
  {journal} {Phys. Rev. A}\ }\textbf {\bibinfo {volume} {56}},\ \bibinfo
  {pages} {131} (\bibinfo {year} {1997})}\BibitemShut {NoStop}%
\bibitem [{\citenamefont {Holevo}(1998)}]{Holevo1998}%
  \BibitemOpen
  \bibfield  {author} {\bibinfo {author} {\bibfnamefont {A.}~\bibnamefont
  {Holevo}},\ }\bibfield  {title} {\bibinfo {title} {The capacity of the
  quantum channel with general signal states},\ }\href
  {https://doi.org/10.1109/18.651037} {\bibfield  {journal} {\bibinfo
  {journal} {IEEE Transactions on Information Theory}\ }\textbf {\bibinfo
  {volume} {44}},\ \bibinfo {pages} {269} (\bibinfo {year} {1998})}\BibitemShut
  {NoStop}%
\bibitem [{\citenamefont {Holevo}\ \emph {et~al.}(1999)\citenamefont {Holevo},
  \citenamefont {Sohma},\ and\ \citenamefont
  {Hirota}}]{Holevo-bosonic-channels99}%
  \BibitemOpen
  \bibfield  {author} {\bibinfo {author} {\bibfnamefont {A.~S.}\ \bibnamefont
  {Holevo}}, \bibinfo {author} {\bibfnamefont {M.}~\bibnamefont {Sohma}},\ and\
  \bibinfo {author} {\bibfnamefont {O.}~\bibnamefont {Hirota}},\ }\bibfield
  {title} {\bibinfo {title} {Capacity of quantum {G}aussian channels},\ }\href
  {https://doi.org/10.1103/PhysRevA.59.1820} {\bibfield  {journal} {\bibinfo
  {journal} {Phys. Rev. A}\ }\textbf {\bibinfo {volume} {59}},\ \bibinfo
  {pages} {1820} (\bibinfo {year} {1999})}\BibitemShut {NoStop}%
\bibitem [{\citenamefont {Holevo}\ and\ \citenamefont
  {Werner}(2001)}]{Holevo-Werner2001}%
  \BibitemOpen
  \bibfield  {author} {\bibinfo {author} {\bibfnamefont {A.~S.}\ \bibnamefont
  {Holevo}}\ and\ \bibinfo {author} {\bibfnamefont {R.~F.}\ \bibnamefont
  {Werner}},\ }\bibfield  {title} {\bibinfo {title} {Evaluating capacities of
  bosonic {G}aussian channels},\ }\href
  {https://doi.org/10.1103/PhysRevA.63.032312} {\bibfield  {journal} {\bibinfo
  {journal} {Phys. Rev. A}\ }\textbf {\bibinfo {volume} {63}},\ \bibinfo
  {pages} {032312} (\bibinfo {year} {2001})}\BibitemShut {NoStop}%
\bibitem [{\citenamefont {Giovannetti}\ \emph {et~al.}(2004)\citenamefont
  {Giovannetti}, \citenamefont {Guha}, \citenamefont {Lloyd}, \citenamefont
  {Maccone},\ and\ \citenamefont {Shapiro}}]{giovannetti2004minimum}%
  \BibitemOpen
  \bibfield  {author} {\bibinfo {author} {\bibfnamefont {V.}~\bibnamefont
  {Giovannetti}}, \bibinfo {author} {\bibfnamefont {S.}~\bibnamefont {Guha}},
  \bibinfo {author} {\bibfnamefont {S.}~\bibnamefont {Lloyd}}, \bibinfo
  {author} {\bibfnamefont {L.}~\bibnamefont {Maccone}},\ and\ \bibinfo {author}
  {\bibfnamefont {J.~H.}\ \bibnamefont {Shapiro}},\ }\bibfield  {title}
  {\bibinfo {title} {Minimum output entropy of bosonic channels: A
  conjecture},\ }\href {https://doi.org/10.1103/PhysRevA.70.032315} {\bibfield
  {journal} {\bibinfo  {journal} {Phys. Rev. A}\ }\textbf {\bibinfo {volume}
  {70}},\ \bibinfo {pages} {032315} (\bibinfo {year} {2004})}\BibitemShut
  {NoStop}%
\bibitem [{\citenamefont {Giovannetti}\ \emph {et~al.}(2014)\citenamefont
  {Giovannetti}, \citenamefont {Garc{\'\i}a-Patr{\'o}n}, \citenamefont {Cerf},\
  and\ \citenamefont {Holevo}}]{giovannetti2014ultimate}%
  \BibitemOpen
  \bibfield  {author} {\bibinfo {author} {\bibfnamefont {V.}~\bibnamefont
  {Giovannetti}}, \bibinfo {author} {\bibfnamefont {R.}~\bibnamefont
  {Garc{\'\i}a-Patr{\'o}n}}, \bibinfo {author} {\bibfnamefont {N.~J.}\
  \bibnamefont {Cerf}},\ and\ \bibinfo {author} {\bibfnamefont {A.~S.}\
  \bibnamefont {Holevo}},\ }\bibfield  {title} {\bibinfo {title} {Ultimate
  classical communication rates of quantum optical channels},\ }\href
  {https://doi.org/10.1038/nphoton.2014.216} {\bibfield  {journal} {\bibinfo
  {journal} {Nature Photonics}\ }\textbf {\bibinfo {volume} {8}},\ \bibinfo
  {pages} {796} (\bibinfo {year} {2014})}\BibitemShut {NoStop}%
\bibitem [{\citenamefont {Mari}\ \emph {et~al.}(2014)\citenamefont {Mari},
  \citenamefont {Giovannetti},\ and\ \citenamefont {Holevo}}]{mari2014quantum}%
  \BibitemOpen
  \bibfield  {author} {\bibinfo {author} {\bibfnamefont {A.}~\bibnamefont
  {Mari}}, \bibinfo {author} {\bibfnamefont {V.}~\bibnamefont {Giovannetti}},\
  and\ \bibinfo {author} {\bibfnamefont {A.~S.}\ \bibnamefont {Holevo}},\
  }\bibfield  {title} {\bibinfo {title} {Quantum state majorization at the
  output of bosonic {G}aussian channels},\ }\href
  {https://doi.org/10.1038/ncomms4826} {\bibfield  {journal} {\bibinfo
  {journal} {Nature Communications}\ }\textbf {\bibinfo {volume} {5}},\
  \bibinfo {pages} {1} (\bibinfo {year} {2014})}\BibitemShut {NoStop}%
\bibitem [{\citenamefont {Guha}(2008)}]{Guha-PhD}%
  \BibitemOpen
  \bibfield  {author} {\bibinfo {author} {\bibfnamefont {S.}~\bibnamefont
  {Guha}},\ }\emph {\bibinfo {title} {Multiple-user quantum information theory
  for optical communication channels}},\ \href@noop {} {Ph.D. thesis},\
  \bibinfo  {school} {Massachusetts Institute of Technology} (\bibinfo {year}
  {2008})\BibitemShut {NoStop}%
\bibitem [{\citenamefont {Garc{\'\i}a-Patr{\'o}n}\ \emph
  {et~al.}(2012)\citenamefont {Garc{\'\i}a-Patr{\'o}n}, \citenamefont
  {Navarrete-Benlloch}, \citenamefont {Lloyd}, \citenamefont {Shapiro},\ and\
  \citenamefont {Cerf}}]{Garcia2012majorization}%
  \BibitemOpen
  \bibfield  {author} {\bibinfo {author} {\bibfnamefont {R.}~\bibnamefont
  {Garc{\'\i}a-Patr{\'o}n}}, \bibinfo {author} {\bibfnamefont {C.}~\bibnamefont
  {Navarrete-Benlloch}}, \bibinfo {author} {\bibfnamefont {S.}~\bibnamefont
  {Lloyd}}, \bibinfo {author} {\bibfnamefont {J.~H.}\ \bibnamefont {Shapiro}},\
  and\ \bibinfo {author} {\bibfnamefont {N.~J.}\ \bibnamefont {Cerf}},\
  }\bibfield  {title} {\bibinfo {title} {Majorization theory approach to the
  {G}aussian channel minimum entropy conjecture},\ }\href
  {https://link.aps.org/doi/10.1103/PhysRevLett.108.110505} {\bibfield
  {journal} {\bibinfo  {journal} {Physical Review Letters}\ }\textbf {\bibinfo
  {volume} {108}},\ \bibinfo {pages} {110505} (\bibinfo {year}
  {2012})}\BibitemShut {NoStop}%
\bibitem [{\citenamefont {Gagatsos}\ \emph {et~al.}(2013)\citenamefont
  {Gagatsos}, \citenamefont {Oreshkov},\ and\ \citenamefont
  {Cerf}}]{gagatsos2013majorization}%
  \BibitemOpen
  \bibfield  {author} {\bibinfo {author} {\bibfnamefont {C.}~\bibnamefont
  {Gagatsos}}, \bibinfo {author} {\bibfnamefont {O.}~\bibnamefont {Oreshkov}},\
  and\ \bibinfo {author} {\bibfnamefont {N.}~\bibnamefont {Cerf}},\ }\bibfield
  {title} {\bibinfo {title} {Majorization relations and entanglement generation
  in a beam splitter},\ }\href
  {https://link.aps.org/doi/10.1103/PhysRevA.87.042307} {\bibfield  {journal}
  {\bibinfo  {journal} {Physical Review A}\ }\textbf {\bibinfo {volume} {87}},\
  \bibinfo {pages} {042307} (\bibinfo {year} {2013})}\BibitemShut {NoStop}%
\bibitem [{\citenamefont {Jabbour}\ and\ \citenamefont
  {Cerf}(2021)}]{jabbour2021multiparticle}%
  \BibitemOpen
  \bibfield  {author} {\bibinfo {author} {\bibfnamefont {M.~G.}\ \bibnamefont
  {Jabbour}}\ and\ \bibinfo {author} {\bibfnamefont {N.~J.}\ \bibnamefont
  {Cerf}},\ }\bibfield  {title} {\bibinfo {title} {Multiparticle quantum
  interference in {B}ogoliubov bosonic transformations},\ }\href
  {https://link.aps.org/doi/10.1103/PhysRevResearch.3.043065} {\bibfield
  {journal} {\bibinfo  {journal} {Physical Review Research}\ }\textbf {\bibinfo
  {volume} {3}},\ \bibinfo {pages} {043065} (\bibinfo {year}
  {2021})}\BibitemShut {NoStop}%
\bibitem [{\citenamefont {Kaftal}\ and\ \citenamefont
  {Weiss}(2010)}]{infinite-majorization}%
  \BibitemOpen
  \bibfield  {author} {\bibinfo {author} {\bibfnamefont {V.}~\bibnamefont
  {Kaftal}}\ and\ \bibinfo {author} {\bibfnamefont {G.}~\bibnamefont {Weiss}},\
  }\bibfield  {title} {\bibinfo {title} {An infinite dimensional
  {S}chur–{H}orn theorem and majorization theory},\ }\href
  {https://doi.org/doi.org/10.1016/j.jfa.2010.08.018} {\bibfield  {journal}
  {\bibinfo  {journal} {Journal of Functional Analysis}\ }\textbf {\bibinfo
  {volume} {259}},\ \bibinfo {pages} {3115} (\bibinfo {year}
  {2010})}\BibitemShut {NoStop}%
\bibitem [{\citenamefont {Jabbour}\ \emph {et~al.}(2016)\citenamefont
  {Jabbour}, \citenamefont {Garc{\'{\i}}a-Patr{\'{o}}n},\ and\ \citenamefont
  {Cerf}}]{Jabbour_2016}%
  \BibitemOpen
  \bibfield  {author} {\bibinfo {author} {\bibfnamefont {M.~G.}\ \bibnamefont
  {Jabbour}}, \bibinfo {author} {\bibfnamefont {R.}~\bibnamefont
  {Garc{\'{\i}}a-Patr{\'{o}}n}},\ and\ \bibinfo {author} {\bibfnamefont
  {N.~J.}\ \bibnamefont {Cerf}},\ }\bibfield  {title} {\bibinfo {title}
  {Majorization preservation of {G}aussian bosonic channels},\ }\href
  {https://doi.org/10.1088/1367-2630/18/7/073047} {\bibfield  {journal}
  {\bibinfo  {journal} {New Journal of Physics}\ }\textbf {\bibinfo {volume}
  {18}},\ \bibinfo {pages} {073047} (\bibinfo {year} {2016})}\BibitemShut
  {NoStop}%
\bibitem [{\citenamefont {De~Palma}\ \emph {et~al.}(2016)\citenamefont
  {De~Palma}, \citenamefont {Trevisan},\ and\ \citenamefont
  {Giovannetti}}]{depalma2016passive}%
  \BibitemOpen
  \bibfield  {author} {\bibinfo {author} {\bibfnamefont {G.}~\bibnamefont
  {De~Palma}}, \bibinfo {author} {\bibfnamefont {D.}~\bibnamefont {Trevisan}},\
  and\ \bibinfo {author} {\bibfnamefont {V.}~\bibnamefont {Giovannetti}},\
  }\bibfield  {title} {\bibinfo {title} {Passive states optimize the output of
  bosonic {G}aussian quantum channels},\ }\href
  {https://doi.org/10.1109/TIT.2016.2547426} {\bibfield  {journal} {\bibinfo
  {journal} {IEEE Transactions on Information Theory}\ }\textbf {\bibinfo
  {volume} {62}},\ \bibinfo {pages} {2895} (\bibinfo {year}
  {2016})}\BibitemShut {NoStop}%
\end{thebibliography}%

\end{document}